# Threats to the information system in the physical environment and cyberspace

Valeria Ageeva[a], Aleksey Novokhrestov[a], Maria Kholodova[a]

[a] Tomsk State University of Control Systems and Radioelectronics, 40 Lenina Prospect, Tomsk 634050, Russia

Correspondence: Aleksey Novokhrestov, nak@fb.tusur.ru

**Abstract**

The purpose of the study is to supplement and update the list of threats to the confidentiality and integrity of the system. The article focuses on the already compiled list of threats and a model of system, but also considers new threats and types of threats. Scientific novelty is in the interdisciplinary consideration of the issue with the involvement of the works of modern Russian and Western scientists. As a result of the study, new threats to the confidentiality and integrity of the system were described, the type of these threats was determined and classified by channels of communication.

**Keywords**: security assessment, information system, model of information system, threat.

**Introduction**

The relevance of threats to the integrity and confidentiality of information requires careful attention to protection against them. The task of ensuring information security is solved by means of cryptographic protection, firewalls, access control and other means of protection. However, now they are no longer enough, as the number of threats is increasing and an integrated approach to their assessment and creation of protection systems is required.

In the course of the analysis of existing and successfully applied models of threats, with the identification of their advantages and problematic aspects, it was concluded that the existing models often contain an incomplete range of information regarding the described threats (for example, [1,2] and [3]). These threat models do not fully describe the threats to the model. Each model describes the same threats, only in different words, or one model complements the other. Also, in these models, threats are described by verbal and mathematical formalizations, which is not very convenient. This can lead to the fact that each threat can be interpreted in its own way.

In this article, an attempt is made to supplement the threat model for the information system, and to formulate the most effective methodology and system of criteria for assessing the level of security of information systems. An attempt was made to combine mathematical and verbal formalizations into a single system.

The uniqueness of the compiled model will lie in its flexibility and versatility. Subsequently, the presented model will contain all currently known threats from various areas (in other words, a bank of threats), as well as countermeasures. In addition, the urgency of the problem being solved lies in the fact that in the Russian Federation, for any information systems that are somehow subject to protection in accordance with the law, it is necessary to develop a threat model.

If we compare the developed model with analogs, then it should become more universal, allowing various enterprises to use it for a comprehensive analysis of their own security system, to identify vulnerabilities and select appropriate countermeasures. The correct use of such a model will practically eliminate the possibility that some element of the system will remain unprotected, that some vulnerability will not be discovered.

**Proposed approach**

This article provides tables with some of the threats to confidentiality and system integrity. The information system consists of territories where buildings with offices are located, where confidential information is stored - this is a physical object of threats. And also from the local area network, workstation (operating system) and software - this is a software threat object.

The developed threat model describes the threats to confidentiality to elements of the information system [1]. The results of the work done for the confidentiality of the system are shown in tables 1 - 3. The tables also categorize threats by threat type. The types of privacy threats are as follows:
Type 1 - disclosure of location information;
Type 2 - disclosure of information about protection mechanisms;
Type 3 - disclosure of information about attributes.

*Table 1 - Territory elements and confidentiality threats*

| Element | List of threats | Types of threats |
|---|---|---|
| Office | Disclosure (leakage) of the office number where confidential information is stored or processed | Type 1 |
| | Disclosure (leakage) of the list of employees who have access to confidential information. Sniffing out information from employees (the one who does not have access to confidential information, who does). Same threats to the building | Type 2 |
| | Disclosure of the level of confidentiality of information that is stored | Type 3 |
| Building | Disclosure of the location of the building and the entrance | Type 1 |
| | Disclosure of the rules of entry for employees | Type 2 |
| | Disclosure of the building plan | Type 3 |

*Table 2 - Territory channels and confidentiality threats*

| Channel | List of threats | Types of threats |
|---|---|---|
| Corridor | Disclosure of the location of the corridors (on which floor, which corridor and which offices are on this floor) | Type 1 |
| | Disclosure of the list of employees who can move along certain corridors and floors | Type 2 |
| | Disclosure of the corridor plan | Type 3 |
| Territory | Disclosure of the plan for the location of buildings on the territory | Type 1 |
| | Disclosure of the list of employees and entry rules for these employees who have access to the territory | Type 2 |
| | Disclosure of the plan of territories | Type 3 |

*Table 3 - Computer network elements and confidentiality threats*

| Element | List of threats | Types of threats |
|---|---|---|
| Software | Disclosure (leakage) of information about the name of the software installed within the operating system | Type 1 |
| | Disclosure (leakage) of information about the protocol by which the software interacts | Type 2 |
| | Disclosure (leakage) of information about the name of the software | Type 3 |
| Operating system (OS) | Disclosure (leakage) of information about OS settings | Type 1 |
| | Disclosure (leakage) of information about the protocol by which the OS interacts in the corresponding local network; | Type 2 |
| | Disclosure (leakage) of information about the OS name. | Type 3 |
| Network (LAN) | Disclosure (leakage) of information about LAN settings; | Type 1 |
| | Disclosure (leakage) of information about the protocol by which the LAN interacts; | Type 2 |
| | Disclosure (leakage) of information about the name of the LAN. | Type 3 |

System integrity threats have also been described. These results are shown in tables 4-5. As well as for threats to the confidentiality of the system, in the tables, threats are distributed by threat type. The types of integrity threats are as follows:

Type 1 - disable or delete;

Type 2 - addition;

Type 3 - substitution;

Type 4 - change of attributes.

*Table 4 - Territory elements, channels, and integrity threats*

| Element / channel | List of threats | Types of threats |
|---|---|---|
| Office | Disable the camera, which is in the office with confidential information | Type 1 |
| | Installation of hidden cameras, microphones or bugs in the office | Type 2 |
| | Substitution of a regular pass for a pass, which makes it possible to enter the office with confidential information | Type 3 |
| | Violation of the efficiency of employees who work in the office | Type 4 |
| Corridor | Disabling or disabling cameras located in the corridors | Type 1 |
| | Introduction of an unauthorized user into the list of employees who can move along the corridors | Type 2 |
| | Replacement of the corridor plan | Type 3 |
| | Changing the building plan | Type 4 |
| Building | Power outages throughout the building | Type 1 |
| | Adding new employees to the list of those employees who can pass through the checkpoint | Type 2 |
| | Steal a pass from employees who have the right to enter | Type 3 |
| | Changing the building plan | Type 4 |

| Territory | Power outages | Type 1 |
|---|---|---|
| | Installing a microphone or bug in the smoking area | Type 2 |
| | Replacement of keys to enter | Type 3 |
| | Distortion of data in the pass, which makes it possible to enter the territory | Type 4 |

*Table 5 - Computer network elements and integrity threats*

| Element | List of threats | Types of threats |
|---|---|---|
| Software | Removal of software | Type 1 |
| | Software installation | Type 2 |
| | Software substitution | Type 3 |
| | Change the port number that the software uses | Type 4 |
| Operating system (OS) | Intentional damage (disabling) of the operating system | Type 1 |
| | Installation of an additional operating system | Type 2 |
| | Substitution of the operating system | Type 3 |
| | Changing the ip-address used by the operating system | Type 4 |
| Network (LAN) | Intentional damage (disablement) of the local computer network | Type 1 |
| | Adding protocols | Type 2 |
| | Substitution of protocols | Type 3 |
| | Changing the ip-address of the network | Type 4 |

**Conclusion**

Work was carried out to supplement the threat model for the information system described in [1-3]. Based on the model of the information system, the types of threats to elements and channels were identified. A large number of threats to the information system have been identified, and for the convenience of their detection, a method for classifying these threats has been compiled. In the course of the work, 6 elements, 6 channels, as well as typical threats were identified: 34 for the confidentiality of the system and 32 for the integrity of the system.

**Conflict of interest**

None.

**Acknowledgements**

This research was funded by the Ministry of Science and Higher Education of Russia, Government Order for 2020–2022, project no. FEWM-2020-0037 (TUSUR).